%% file: main_arxiv.tex
\documentclass{article}

 \usepackage[preprint]{main_arxiv}

\usepackage[utf8]{inputenc} 
\usepackage[T1]{fontenc}    

\usepackage{upquote}

\usepackage{hyperref}       
\usepackage{url}            
\usepackage{booktabs}       
\usepackage{amsfonts}       
\usepackage{nicefrac}       
\usepackage{microtype}      
\usepackage{xcolor}         

\usepackage{graphicx}
\usepackage{multirow}
\usepackage{makecell}
\usepackage{amsmath}
\usepackage[table]{xcolor}
\usepackage{tikz}
\usetikzlibrary{calc}
\usepackage[most]{tcolorbox}
\usepackage[most]{mathtools}
\usepackage{caption}

\usepackage{amsthm}
\theoremstyle{definition}
\newtheorem{definition}{Definition}

\usepackage{wrapfig}

\definecolor{myblue1}{RGB}{218, 232, 252}
\definecolor{myblue2}{RGB}{108, 142, 191}
\definecolor{myblue3}{RGB}{153, 204, 255}

\newcommand{\Tool}{\textsc{CodePivot}}

\newcommand{\Dataset}{$\mathcal{D}_\textsc{Py-Trans}$}

\newcommand{\Datasetnm}{\mathcal{D}_\textsc{Py-Trans}}

\newcommand{\Distilldataset}{$\mathcal{D}_\textsc{Py2Others-SFT}$}

\newcommand{\RLdataset}{$\mathcal{D}_\textsc{Py2Others-RL}$}

\newcommand{\Benchmark}{$\mathcal{D}_\textsc{Py2OthersBench}$}

\newcommand{\Benchmarkothers}{$\mathcal{D}_\textsc{Others2AllBench}$}

\newcommand{\Powerfulllm}{Claude-Sonnet-4}

\newcommand{\Basemodel}{Qwen2.5-Coder-7B-Instruct}

\newcommand{\Program}{p}
\newcommand{\Prog}[1]{p_\mathrm{#1}}

\newcommand{\Lang}[1]{l_\mathrm{#1}}

\newcommand{\Testset}{\tau}

\newcommand{\Prompt}{q}

\newcommand{\FullTargetLangsList}{\textsf{C++}, \textsf{C\#}, \textsf{Java}, \textsf{JavaScript}, \textsf{Golang}, \textsf{Perl}, \textsf{Ruby}, \textsf{Rust}, and \textsf{Haskell}}

\newcommand{\Base}{Base-7B}
\newcommand{\BaseSft}{Base-7B-SFT}

\newcommand{\RLLinear}{$\text{Base-7B-RL-}\mathcal{R}_\text{linear}$}
\newcommand{\RLFlip}{$\text{Base-7B-RL-}\mathcal{R}_\text{conservative}$}
\newcommand{\RLDiscrete}{$\text{Base-7B-RL-}\mathcal{R}_\text{discrete}$}
\newcommand{\RLPartial}{$\text{Base-7B-RL-}\mathcal{R}_\text{aggressive}$}

\newcommand{\RewardName}{\emph{Aggressive-Partial-Functional} reward}

\newcommand{\LangGroup}[1]{\makecell[c]{\textbf{#1}}}

\newcommand{\autotabref}[1]{Table~\ref{#1}}

\title{\Tool{}: Bootstrapping Multilingual Transpilation in LLMs via Reinforcement Learning without Parallel Corpora}

\author{%
  Shangyu Li$^{1}$\quad
  Juyong Jiang$^{2}$\quad
  Meibo Ren$^{1}$\quad
  Sizhe Zhong$^{1}$\quad
  Huiri Tan$^{1}$\quad
  Yunhao Gou$^{1}$\\
  \textbf{Xu Han}$^{2}$\quad
  \textbf{Chun Yong Chong}$^{3}$\quad
  \textbf{Yun Peng}$^{4}$\quad
  \textbf{Jiasi Shen}$^{1}$\\[6pt]
  $^{1}$The Hong Kong University of Science and Technology \\
  $^{2}$The Hong Kong University of Science and Technology (Guangzhou)\\
  $^{3}$Monash University\quad
  $^{4}$The Chinese University of Hong Kong\\[4pt]
  \texttt{\{sliew, mrenad, sizhe.zhong, htanaj, ygou, sjs\}@connect.ust.hk}\\
  \texttt{\{jjiang472, xhan696\}@connect.hkust-gz.edu.cn}\\
  \texttt{Chong.chunyong@monash.edu\quad ypeng@cse.cuhk.edu.hk}\\[4pt]
}

\begin{document}

\maketitle

\begin{abstract}
Transpilation, or code translation, aims to convert source code from one programming language (PL) to another. It is beneficial for many downstream applications, from modernizing large legacy codebases to augmenting data for low-resource PLs.
Recent large language model (LLM)-based approaches have demonstrated immense potential for code translation.
Among these approaches, training-based methods are particularly important because LLMs currently do not effectively adapt to domain-specific settings that suffer from a lack of knowledge without targeted training. This limitation is evident in transpilation tasks involving low-resource PLs.
However, existing training-based approaches rely on a pairwise transpilation paradigm, making it impractical to support a diverse range of PLs. This limitation is particularly prominent for low-resource PLs due to a scarcity of training data. Furthermore, these methods suffer from suboptimal reinforcement learning (RL) reward formulations.
To address these limitations, we propose \Tool{}, a training framework that leverages Python as an intermediate representation (IR), augmented by a novel RL reward mechanism, \RewardName{}, to bootstrap the model's multilingual transpilation ability without requiring parallel corpora.
Experiments involving 10 PLs show that the resulting 7B model, trained on Python-to-Others tasks, consistently improves performance across both general and low-resource PL-related transpilation tasks.
It outperforms substantially larger mainstream models with hundreds of billions more parameters, such as Deepseek-R1 and Qwen3-235B-A22B-Instruct-2507, on Python-to-Others tasks and Others-to-All tasks, respectively. In addition, it outperforms its counterpart trained directly on Any-to-Any tasks on general transpilation tasks.
The code and data are available at \href{https://github.com/lishangyu-hkust/CodePivot}{https://github.com/lishangyu-hkust/CodePivot}.
\end{abstract}

\input{section/new_intro}

\input{section/related_work}

\input{section/design}

\input{section/implementation}

\input{section/evaluation}

\input{section/discussion}

\input{section/conclusion}

\bibliographystyle{plain}
\bibliography{reference}




\appendix

\input{section/appendix}


\end{document}

%% file: section/new_intro.tex
\section{Introduction}

Transpilation (code translation), which involves converting source code from one programming language (PL) to another, is particularly important in practice.
It supports key practical needs such as migrating legacy systems to modern platforms~\cite{sneed2010migrating, ziftci2025migrating}, reducing long-term maintenance costs~\cite{sneed2010migrating}, strengthening security by moving from unsafe PLs to safer ones~\cite{emre2021translating, ling2022rust, hans2025automated, sneed2010migrating, hong2025type, hong2024tag, nitin2025c2saferrust}, and creating parallel corpora that improve code-synthesis training and mitigate data scarcity for low-resource PLs~\cite{pan2024lost}.

Despite decades of research, high-quality transpilation remains challenging. Classical rule-based approaches~\cite{wang2023user} depend on heavy static analysis and extensive engineering, and they often fail when language semantics and idioms do not align. Program synthesis-based methods~\cite{mariano2022automated, 10.1109/52.895180} 
tend to scale poorly due to search complexity and the cost of validation.

Recent large language models (LLMs) offer real-time transpilation~\cite{intertrans, yang2024exploring, bhatia2024verified, yang2024vert, ibrahimzada2025alphatrans, feischl2025large} and may alleviate the scalability limitations of synthesis~\cite{zhang2025entireprojects, wang2024repotransbench, yuan2025project}.
Among these approaches, training-based methods are particularly important for addressing this problem because, without targeted training, LLMs currently do not effectively adapt to domain-specific settings that suffer from a lack of knowledge~\cite{cassano2024knowledge, gururangan-etal-2020-dont}.
This limitation is particularly evident in low-resource PLs~\cite{wang2025translating} and in tasks such as migrating COBOL to Java~\cite{hans2025automated, sneed2010migrating}, where COBOL remains critical because many financial systems depend on it, yet it is difficult to maintain and modernize due to its decades-old design and the limited number of developers with relevant expertise.
Furthermore, with the emergence of tools such as GitHub Copilot~\cite{Microsoft2026Copilot} and models such as OpenAI Codex~\cite{chen2021evaluating}, developer productivity has substantially improved through automation of repetitive tasks, real-time code suggestions, and detailed explanations of code behavior. However, limited support for low-resource languages places their programmers at a significant disadvantage~\cite{zheng2023codegeex, zhang2024bridge}, as they do not receive the same benefits that code LLMs provide for high-resource languages~\cite{ziegler2022productivity, murali2024ai}.

\input{figures/low_resource_perf}

As illustrated in \autoref{fig:low_resource_perf}, a model's overall coding proficiency exhibits a strong positive correlation with the resource availability of the target programming language. This dependency underscores a critical limitation of current models, which consistently suffer from degraded performance when applied to low-resource languages.

However, existing training-based approaches rely on a pairwise transpilation paradigm, making it impractical to support a wide variety of PLs.
Since these methods specifically require formulating $N(N-1)$ directed transpilation pairs to support $N$ languages~\cite{wang2025effireasontrans, jana2024cotran}, they require $O(N^2)$ language pairs for training, making it difficult to construct sufficient data to enhance the models’ transpilation ability in practice.
This challenge is further exacerbated by the limited availability of high-quality parallel data for low-resource PLs.
Furthermore, while previous work~\cite{wang2025effireasontrans, jana2024cotran} has demonstrated the potential of reinforcement learning (RL) in transpilation, current methods are hindered by suboptimal reward formulations. By relying heavily on either symbolic compiler feedback~\cite{jana2024cotran} or sparse, execution-based rewards derived from the absolute number of test cases passed, existing approaches have yet to effectively leverage RL to improve model performance in transpilation.

To close these gaps, we propose \Tool{}, a training framework that employs Python as an intermediate representation (IR) along with a novel reward mechanism, \RewardName{}, to bootstrap the model's multilingual transpilation abilities without requiring parallel corpora and to harness the potential of RL.
To solve a transpilation task from a PL $\Lang{src}$ to another PL $\Lang{tar}$, we train a model using transpilation data that consists of Python-to-$\Lang{src}$ and Python-to-$\Lang{tar}$ tasks. In other words, we introduce the Python language as an IR or a shared representation \cite{johnson-etal-2017-googles} in the training process to address the original transpilation task from $\Lang{src}$ to $\Lang{tar}$.
Based on Python, \Tool{} implicitly aligns the underlying logic of diverse PLs within the model's latent space, thereby enabling zero-shot transfer across unseen transpilation directions.

The motivation for using Python as an IR stems from prior work~\cite{metacompilertranslation_szafraniec2022code}, which demonstrated that leveraging a low-level compiler IR, such as LLVM IR~\cite{lattner2004llvm}, can facilitate transpilation.
However, this approach is fundamentally constrained to PLs that can be compiled to LLVM IR, such as C, C++, Java, and Rust.
Consequently, it lacks comprehensive multilingual extensibility across a broader spectrum of PLs and remains limited by the inherent data scarcity of low-resource PLs.
It also requires massive datasets of program-to-IR pairs simply for the model to acquire a foundational understanding of the low-level IR~\cite{paul-etal-2024-ircoder, metacompilertranslation_szafraniec2022code}, rendering it highly vulnerable to data scarcity when applied to low-resource PLs.
These limitations motivate the idea that, rather than training models to understand low-level compiler IRs that current LLMs often do not master, it would be more effective to adopt a high-resource PL as the IR.
We therefore select Python as the IR to bridge diverse PLs.
Python is widely recognized as a high-resource PL~\cite{yang2025scaling, multipl-e}, underpinned by an abundance of accessible, high-quality training corpora~\cite{lozhkov2024starcoderstackv2, benallal2024smollmcorpus}.
Additionally, its expressive, natural language-like syntax aligns intuitively with the semantic comprehension capabilities of LLMs~\cite{pandey2024towards}.
By using Python to abstract the core functional behavior of programs, our approach eliminates tedious and error-prone manual engineering effort to model special language features, such as Rust's ownership rules or pointer-level memory semantics.
Consequently, \Tool{} substantially reduces reliance on scarce direct parallel corpora and extends robust transpilation support to 10 PLs, including low-resource PLs.
Under this design, we apply a Python-IR-pivoted supervised fine-tuning (SFT) strategy to bootstrap the LLMs' multilingual transpilation capabilities using exclusively Python-to-Others transpilation pairs without parallel corpora.

To further enhance performance during the RL phase, our newly proposed \RewardName{} emphasizes rollouts with programs that pass at least one but not all test cases (partially functionally correct), encouraging the model to turn more rollouts with programs that pass no test cases (functionally incorrect) into rollouts with partially functionally correct ones.
This mechanism does not rely on heavyweight compiler symbolic feedback~\cite{jana2024cotran} and provides a denser, more informative feedback signal for the policy model compared to conventional rewards based purely on the absolute number of passed test cases~\cite{wang2025effireasontrans}.

We build a file-level transpilation dataset, \Dataset{}, based on PyEdu-R~\cite{li2025codeio, benallal2024smollmcorpus, lozhkov2024starcoderstackv2}.
\Dataset{} contains approximately 1 million Python-to-Others transpilation tasks, where the source PL is Python and the target PLs encompass 9 languages spanning different programming paradigms, compilation models, type systems, and resource abundance, including \FullTargetLangsList{}.
Each transpilation task in \Dataset{} contains a Python source program, language-agnostic test cases, and corresponding metadata.
The Python source programs of the transpilation tasks span a broad array of computational domains, such as algorithms, logic puzzles, math, scientific computation, and system modeling.
To reflect real-world code migration scenarios, the selected Python programs also invoke external APIs, including container management and file I/O operations.

To evaluate the multilingual transpilation performance of the models, we sample a subset of 900 unique Python-to-Others tasks from \Dataset{} to construct a Python-to-Others transpilation benchmark. The tasks are evenly distributed across the 9 target PLs. We sample these tasks without replacement and curate the selection to promote problem variety and structural diversity.
We further sample a subset of Python-to-Others tasks without replacement from \Dataset{} to distill \Powerfulllm{} and obtain the corresponding translated programs. We then filter the translated programs by execution results, yielding 100 oracle programs for each of the 9 target PLs.
These oracle programs subsequently serve as the source programs for the Others-to-All benchmark.
By pairing these source programs across the 9 PLs with 10 target PLs, which include Python, we obtain a comprehensive suite of 8.1K unique transpilation tasks.
In contrast to existing benchmarks that are restricted to function- or class-level transpilations~\cite{multipl-e, xue2025classeval}, feature a narrow and homogeneous set of PLs~\cite{ahmad2023avatar, Unsupervised_translation, zheng2023codegeex}, or lack overall problem diversity~\cite{ahmad2023avatar, Unsupervised_translation}, our proposed benchmarks offer file-level transpilation, multilingual scope, and more extensive problem diversity.

For training purposes, we use a refined subset of fewer than 30K tasks sampled from \Dataset{} without replacement due to computational limitations.
We exclude all benchmark instances from this subset to maintain evaluation integrity.
Using this subset, we instantiate our approach with a 7B-parameter model, \Basemodel{}, by training it exclusively on Python-to-Others transpilation tasks.

Experiments show that the resulting model consistently improves performance across both general and low-resource PL-related transpilation tasks.
Remarkably, it outperforms substantially larger mainstream models with hundreds of billions more parameters, such as Qwen3-235B-A22B-Instruct-2507 and DeepSeek-R1.
Specifically, the resulting model yields performance improvements of 10.67\% on low-resource PL-related tasks and 15.19\% on Others-to-All transpilation tasks relative to Qwen3-235B-A22B-Instruct-2507, while simultaneously outperforming DeepSeek-R1 by 5.34\% on Python-to-Others transpilation tasks.
In addition, experimental results show that the model trained on Python-to-Others tasks is more effective in transpilation than its counterpart trained directly on Any-to-Any tasks.

In summary, our contributions are as follows:
\begin{itemize}
    \item To the best of our knowledge, we are the first to propose a training framework using Python as an intermediate representation to bootstrap the model's multilingual transpilation capabilities without requiring parallel corpora. Our approach achieves multilingual transpilation training of as many as 10 PLs. Experiments show that it works well even with low-resource PLs whose training data is scarce.
    \item We propose a novel RL reward, \RewardName{}, to enable our framework to improve models' multilingual transpilation capabilities. Experiments show that by incorporating the reward for training \Basemodel{}, our framework yields a 7B-parameter model that surpasses DeepSeek-R1 (671B) in our experiments.
    \item We conduct comprehensive experiments showing the effectiveness of \Tool{} in Others-to-All and Python-to-Others transpilation tasks, in which it surpasses models with hundreds of billions more parameters, and the efficiency of Python as an IR, which enables it to surpass its counterpart trained on Any-to-Any tasks by 5.83\%.
\end{itemize}

%% file: figures/low_resource_perf.tex
\begin{wrapfigure}{r}{0.6\textwidth}
    \centering
    \includegraphics[width=\linewidth]{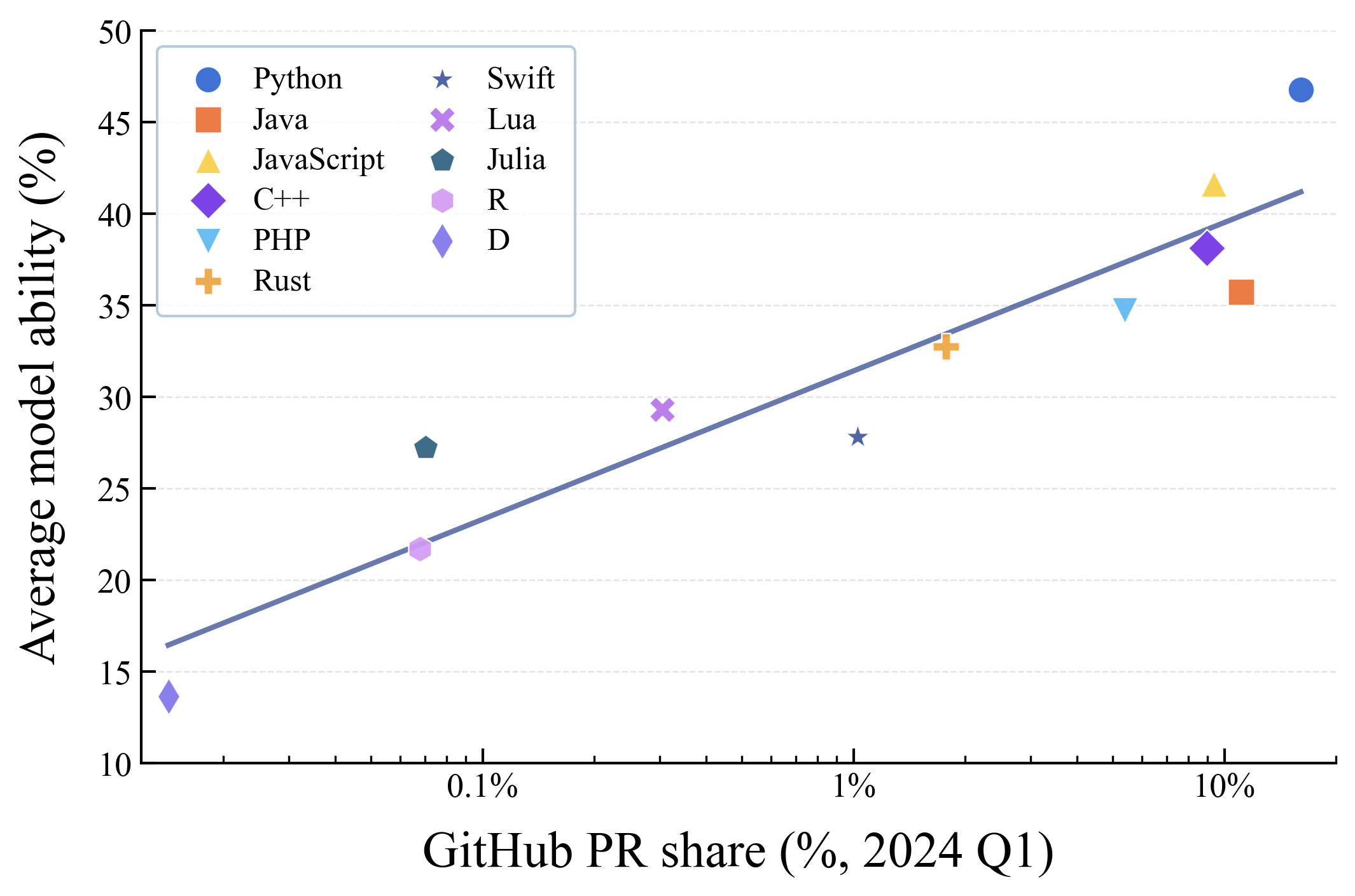}
    \caption{Correlation between programming language resource abundance and average model coding proficiency. The y-axis denotes the coding capability scores averaged across models evaluated by the Big Code Models Leaderboard~\cite{bigcode_models_leaderboard, li2023starcoder}, while the x-axis (logarithmic scale) represents the resource availability of each language, quantified by its respective percentage of GitHub pull requests in 2024 Q1~\cite{githut2}. A logarithmic trendline is fitted to highlight the positive correlation.
    }
    \label{fig:low_resource_perf}
\end{wrapfigure}

%% file: section/related_work.tex
\section{Related work}

\paragraph{Code translation}

Transpilation, or code translation, has been studied for a long time~\cite{xue2025classeval} and has motivated extensive research.
Early efforts primarily relied on rule-based methods~\cite{wang2023user, sharpen, c2rust, cxgo, j2cstranslator} that require hand-crafted rules for each translation setting, which limit scalability and make it difficult to generalize across languages and complex programming patterns.
More recently, learning-based and LLM-based approaches have become an active direction~\cite{xue2025classeval, pan2024lost, bhattarai2024enhancing, ou2024repository, ramos2024batfix, multipl-e}.
Prior work strengthens the core code translation capability of LLMs by leveraging large-scale data and improved training algorithms \cite{puri2021codenet, zheng2023codegeex, Unsupervised_translation, paul-etal-2024-ircoder, metacompilertranslation_szafraniec2022code, cassano2024knowledge}, while other studies rely on prompting and pipeline orchestration to handle challenging cases~\cite{intertrans, yang2024exploring}. Although these methods often generalize well, they typically provide weaker guarantees of correctness.
Some approaches investigate the synergy between rule-based methods and LLMs to balance correctness with generalization~\cite{ibrahimzada2025alphatrans, bhatia2024verified, yang2024vert}.

\paragraph{Reinforcement learning for code.}
Reinforcement learning has become an effective approach for improving the reasoning capabilities of large language models~\cite{guo2024deepseek, grposhao2024deepseekmath, li2022competition, wang2024enhancing}, with particularly notable gains in code- and mathematics-oriented domains.
Recent studies have applied reinforcement learning to enhance code generation, including using execution feedback for code synthesis to improve generation quality~\cite{gehring2024rlef, shojaee2023execution}, training for Verilog program generation~\cite{zhu2025codev}, and leveraging synthesized test cases to further strengthen code generation~\cite{zeng2025acecoder}.
In addition to code generation, reinforcement learning has also been explored for broader programming tasks, including issue resolution~\cite{wei2025swe}, code translation~\cite{jana2024cotran, wang2025effireasontrans}, reasoning and planning~\cite{rlreason2025ay} 
, and source code summarization~\cite{wang2020reinforcement}.

\paragraph{Intermediate representation.}
An intermediate representation is an abstract form of a program and has been a foundational concept since the early development of compiler design~\cite{strong1958problem, huskey1960neliac, merrill2003generic, davidson1980design, lattner2004llvm}.
LLVM IR stands out for its language agnosticism, well-defined semantics, and rich toolchain support, which have made it a widely adopted foundation for modern research in static analysis for bug detection~\cite{sui2016demand, shi2018pinpoint, sui2016svf, sui2012static, sui2014detecting, tang2024octopus, chen2025seal}, program synthesis and migration~\cite{zhang2024siro, li2025sprout}, verification~\cite{nelson2017hyperkernel, lopes2021alive2, li2025osvbench}, and compiler optimization~\cite{lopes2015provably, tan2025profix}.
More recently, LLVM IR has also been used to study and enhance the multilingual capabilities of code models. Prior work~\cite{trainllm2comprehendllvmir} trains models to learn compiler IR, while other studies~\cite{metacompilertranslation_szafraniec2022code} leverage LLVM IR to improve transpilation performance. In addition, some studies~\cite{paul-etal-2024-ircoder} investigate the use of compiler intermediate representations to improve the multilingual capabilities of Code LMs and facilitate cross-lingual transfer.

%% file: section/design.tex
\section{Methodology}

This section establishes the foundational concepts and outlines the architectural design of \Tool{}.

\subsection{Problem Formulation}
\Tool{} focuses on the source-to-source transpilation task of translating a program from a source PL to a target PL. Next, we present the foundational concepts of the problem.

\begin{definition}[Test Cases]
A \emph{test case} $(s_i, s_o)$ is a pair of input and output strings.
\end{definition}

\begin{definition}[Programs]
A \emph{program} $p$ is modeled as a function that accepts an input string $s_i$ and returns an output string $s_o$; i.e., the function is denoted as $p(s_i) = s_o$.
\end{definition}

In this paper, we focus exclusively on deterministic programs.
We exclude nondeterministic programs since they can produce different outputs for the same input, leading to unpredictable behavior during evaluation.

\begin{definition}[Functional Correctness]
Given a set of test cases $\tau$ and a program $p$, the program $p$ is said to be \emph{functionally correct} iff $p$ passes all the test cases in $\tau$.
The program $p$ is said to be \emph{partially functionally correct} iff $p$ passes at least one test case in $\tau$ but not all the test cases in $\tau$.
The program $p$ is said to be \emph{functionally incorrect} iff $p$ passes no test cases in $\tau$.
\end{definition}

\begin{definition}[Behavioral Equivalence]
Given a set of test cases $\Testset$ and two programs $p_1$ and $p_2$, $p_1$ is \emph{behaviorally equivalent} to $p_2$ modulo $\Testset$ iff $p_1$ and $p_2$ pass the same subset of test cases in $\Testset$.
\end{definition}

\begin{definition}[Transpilation Task]
A \emph{transpilation task} is a tuple $(\Prog{src}, \Testset, \Lang{src}, \Lang{tar})$, where $\Prog{src}$ is the source program written in the source PL $\Lang{src}$, $\Testset$ is a finite set of test cases specifying the intended input-output behavior of $\Prog{src}$, and $\Lang{tar}$ is the target PL.

A target program $\Prog{tar}$ is a \emph{valid} result of the transpilation task iff $\Prog{tar}$ is written in $\Lang{tar}$ and $\Prog{tar}$ is behaviorally equivalent to $\Prog{src}$ modulo $\Testset$.

\end{definition}

\input{figures/workflow-training}

\subsection{Python-IR-pivoted Supervised Fine-Tuning}
\label{sec:python-ir-sft}

In addition to the inspiration drawn from prior work~\cite{metacompilertranslation_szafraniec2022code}, we propose this IR-based training approach guided by hypothetico-deductive reasoning.

Given a set of PLs $L$, we introduce an IR as a shared semantic representation~\cite{johnson-etal-2017-googles} across the PLs in $L$ and hypothesize that for any $\Lang{s}, \Lang{t} \in L$, improvements in both $\Lang{s}$-to-IR and IR-to-$\Lang{t}$ tasks enhance $\Lang{s}\text{-to-}\Lang{t}$ transpilation performance.
In addition, training on $\Lang{s}$-to-$\Lang{t}$ tasks teaches the model the relationship between $\Lang{s}$ and $\Lang{t}$ and improves transpilation performance. We hypothesize that these improvements enhance performance on $\Lang{t}$-to-$\Lang{s}$ tasks.

Under these assumptions, training on IR-to-$l$ tasks for all PLs $l \in L$ may enable the model to learn relationships among the PLs in $L$ and improve performance across tasks.
We hypothesize that these improvements enhance $l$-to-IR performance and further strengthen $\Lang{s}$-to-$\Lang{t}$ transpilation.
In this setting, without an IR, improving general transpilation among PLs in $L$ requires constructing \(\lvert L \rvert(\lvert L \rvert-1)\) PL-to-PL training pairs.
With an IR bridge, training requires only \(\lvert L \rvert\) IR-to-PL pairs, which reduces the number of language pairs from \(O(\lvert L \rvert^2)\)~\cite{wang2025effireasontrans, jana2024cotran} to \(O(\lvert L \rvert)\) and alleviates the need for extensive parallel corpora for low-resource PL pairs.

Since our primary focus is on the functional correctness of translated programs, we only use the IR to abstract core functional behaviors rather than model language-specific features, such as Rust’s ownership rules or pointer-level memory semantics. Consequently, we adopt Python as the IR in \Tool{}. As a high-resource PL~\cite{multipl-e}, Python features an expressive, natural language-like syntax that intuitively aligns with the semantic processing capabilities of LLMs~\cite{pandey2024towards}, thereby facilitating robust code comprehension.

Leveraging Python, we propose a Python-IR-pivoted supervised fine-tuning (SFT) approach to bootstrap the model's general transpilation capability.
As illustrated in \autoref{fig:workflow-training}, the process begins by sampling a subset of Python-to-Others transpilation tasks without replacement from \Dataset{}, and this subset is subsequently used to distill an LLM, \Powerfulllm{}.
To ensure data correctness, we employ execution-based rejection sampling, retaining exclusively those responses that yield a functionally correct target program, \(\Prog{tar}\).
Each validated response is then coupled with its corresponding Python-to-Others transpilation prompt to form a distillation pair, systematically populating our SFT dataset, \Distilldataset{}.
Through the iterative application of this curation pipeline, we construct \Distilldataset{} to encompass approximately 2K high-quality distillation pairs for each target PL.
Subsequently, \Tool{} leverages Python as an IR to perform SFT strictly on these Python-to-Others transpilation pairs.
Consequently, this methodology bootstraps the model's general transpilation capability while mitigating the need to train on extensive parallel corpora for every language combination.

\subsection{Aggressive-Partial-Functional Reward Augmented Reinforcement Learning}
\label{sec:rl-stage}

During the \RewardName{}-augmented RL stage, as illustrated in \autoref{fig:workflow-training}, for each target PL, we first sample without replacement 800 Python-to-Others transpilation tasks that are different from the tasks in \Distilldataset{}, thereby constructing the RL training set \RLdataset{}.

Given a transpilation task $(\Prog{src}, \tau, \Lang{src}, \Lang{tar})$, let $q$ denote the natural language prompt that describes the transpilation task, and let $\pi_\theta(\cdot \mid q)$ denote a conditional text-generation model.
We say that the transpilation task is successfully solved by $\pi_\theta(\cdot \mid q)$ if and only if there exists a generated string \(o \sim \pi_\theta(\,\cdot \mid \Prompt)\) such that $o$ contains a program $\Prog{tar}$ that is written in $\Lang{tar}$ and is functionally equivalent to $\Prog{src}$.

During RL, we train the policy LLM $\pi_\theta(\cdot \mid q)$ to solve transpilation tasks, and after $\pi_\theta(\cdot \mid q)$ generates a rollout $o$, we compute the reward used for optimization.
Execution-based rewards are commonly adopted in coding-related RL~\cite{wang2025effireasontrans}.
However, conventional linear execution-based reward schemes assign rewards only based on the number of test cases passed and ignore the underlying states of programs implied by the number of passed test cases.
Given a rollout $o$ that contains a target program $\Prog{tar}$, we classify $\Prog{tar}$ into three states: functionally incorrect, partially functionally correct, and functionally correct.
Specifically, if $\Prog{tar}$ is functionally incorrect, it indicates that $\Prog{tar}$ lacks the essential main logic.
If $\Prog{tar}$ is partially functionally correct, it indicates that $\Prog{tar}$ largely implements the main logic but fails to handle some overlooked corner cases.
If $\Prog{tar}$ is functionally correct, it implements the main logic correctly and passes all corner cases covered by the given test cases.
We hypothesize that a functionally incorrect program cannot easily become partially functionally correct because it does not yet possess nearly correct main logic, whereas a partially functionally correct program can more readily become functionally correct because it already has nearly correct main logic and has only a few corner cases to resolve.

Based on this intuition, we propose the \emph{Aggressive-Partial-Functional reward} to assign a higher reward weight to rollouts with programs that are partially functionally correct.

\begin{definition}[Aggressive-Partial-Functional Reward]
This reward is defined as:
\begin{align}
\label{eq:reward}
\mathcal{R}(o) = \sigma \left ( \frac{1}{\vert \Testset{} \vert}  \Bigl \vert \{ (s_i, s_o) \in \Testset{} \mid \Prog{o}(s_i) = s_o \} \Bigr \vert \right ) + \mathcal{R}_0,
\end{align}
where $\Program{}_o$ is extracted from $o$, and $\mathcal{R}_{\mathrm{0}}$ denotes the reward assigned to rollouts that comply with the required format specified in the prompt $\Prompt$.
Furthermore, $\sigma$, which is a concave function, is the gate function of \RewardName{} and is defined as follows:
\begin{align}
\label{eq:reward-gate}
\sigma(x) = \frac{1-\lambda^{-x}}{1-\lambda^{-1}}, \text{ such that } \lambda > 1,
\end{align}
where $\lambda$ is a hyperparameter that controls the extent to which we reward partial functional correctness.
\end{definition}

In this way, we can encourage the policy model to turn more rollouts with functionally incorrect programs into rollouts with partially functionally correct ones.

Thus, when applying the standard GRPO algorithm~\cite{guo2024deepseek} for RL training, the policy LLM $\pi_\theta(\cdot \mid \Prompt)$ produces multiple rollouts $o$ for each input prompt $\Prompt$. Given the group size $G$, the rollouts are denoted by $\{o_i\}_{i=1}^G \sim {\pi_{\theta_{\mathrm{old}}}}(\cdot \mid \Prompt)$.
Therefore, the policy LLM aims to optimize the following GRPO objective:
\begin{equation}
\begin{aligned}
\mathcal{J}(\theta) = \mathbb{E} \bigg[ \frac{1}{G} \sum_{i=1}^{G} & \frac{1}{|o_i|}\sum_{t=1}^{|o_i|} \bigg\{ \min \bigg[ \frac{\pi_{\theta}(o_{i, t} \mid q, o_{i, <t})}{\pi_{{\theta}_{\mathrm{old}}}(o_{i, t} \mid q, o_{i, <t})} \hat{A}_{i, t}, \\
& \text{clip} \left( \frac{\pi_\theta(o_{i, t} \mid q, o_{i, <t})}{\pi_{{\theta}_{\mathrm{old}}}(o_{i, t} \mid q, o_{i, <t})}, 1 - \epsilon, 1 + \epsilon \right) \hat{A}_{i, t} \bigg] \\ 
& - \beta D_{KL}(\pi_{\theta}(\cdot \mid q) \parallel \pi_{\text{ref}}(\cdot \mid q)) \bigg\} \bigg]
\end{aligned}
\end{equation}
where $\epsilon$ and $\beta$ are hyperparameters, and $\pi_\theta(o_i \mid q)$ and $\pi_\text{ref}(\cdot \mid \Prompt)$ are the probability distributions of the policy and reference models, respectively.
The reward $r_i = \mathcal{R}(o_i)$ and the advantages $\hat{A}_{i, t} = \frac{r_i - \text{mean}(r_1, ..., r_G)}{\text{std}(r_1, ..., r_G)}$ are calculated using the normalized rewards within each group following GRPO. $D_{KL}$ denotes the estimated KL-divergence~\cite{schulman2020kl}.

%% file: figures/workflow-training.tex
\begin{figure*}[t]
    \centering
    \includegraphics[width=\linewidth]{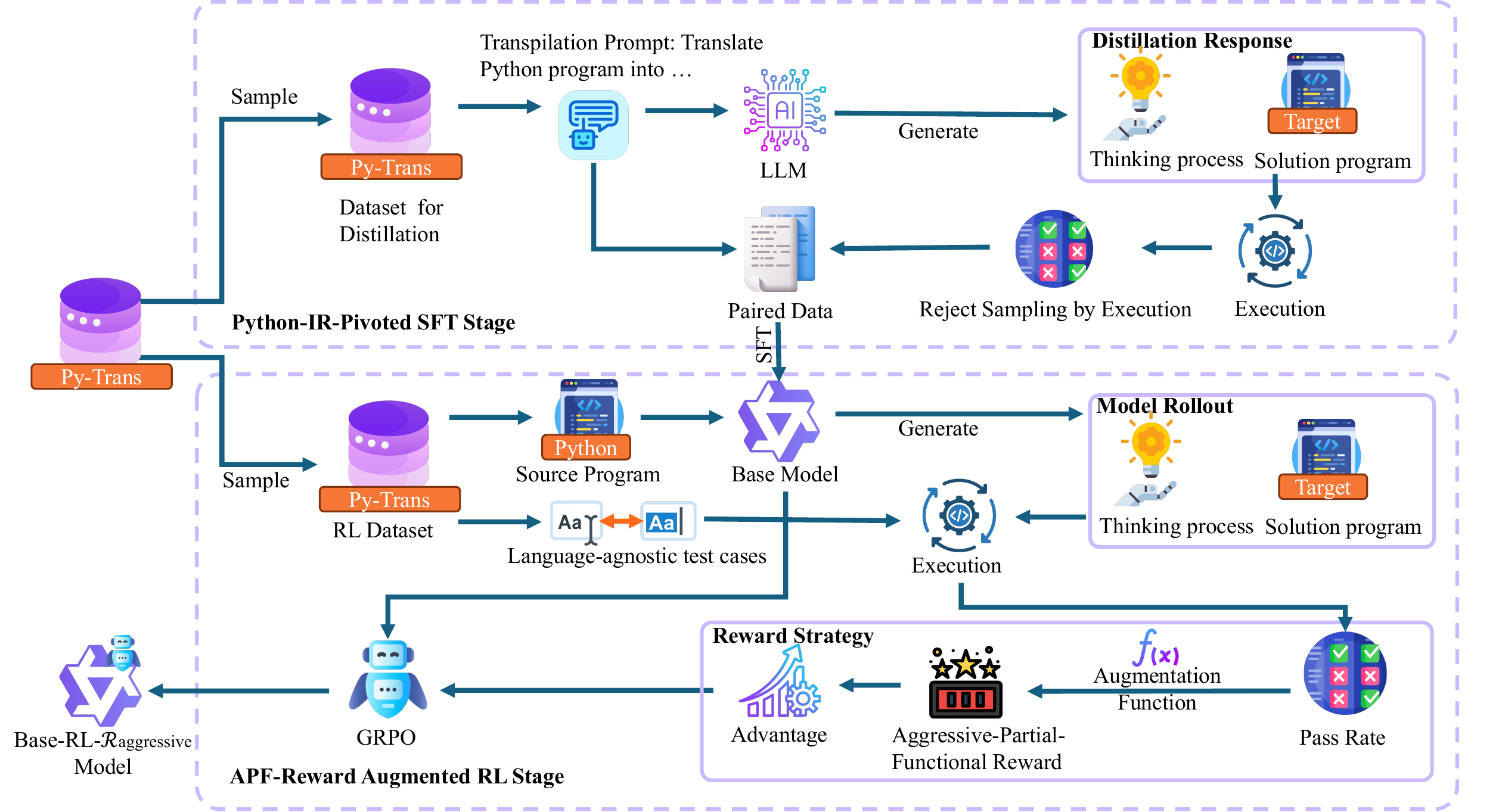}
    \caption{
    The workflow of the training process of \Tool{}. The core components of the workflow consist of the Python-IR-pivoted SFT stage (Section~\ref{sec:python-ir-sft}) and the \RewardName{} (APF) augmented RL stage (Section~\ref{sec:rl-stage}).
}
    \label{fig:workflow-training}
\end{figure*}

%% file: section/implementation.tex
\section{Implementation}

During the Python-IR pivoted supervised fine-tuning (SFT) stage, we use LLaMAFactory \cite{zheng2024llamafactory} to perform SFT of the open-source base model Qwen2.5-Coder-7B-Instruct on \Distilldataset{}. We train the model for 3 epochs with a learning rate of $1 \times 10^{-5}$ and a batch size of 32. The distilled model is used for later RL.

During the RL stage, we implement \Tool{} with a non-linear \RewardName{} in the verl~\cite{verl} framework and further train the distilled model on \RLdataset{} using GRPO~\cite{grposhao2024deepseekmath}. We set the total context length to 22,528, train with a batch size of 32, a clipping range $\epsilon$ of 0.2, and a learning rate of $1 \times 10^{-6}$, and run for about 350 steps until the reward curves converge. The rollout temperature is fixed at 1.0, and the maximum lengths are set to 5,120 tokens for the instruction and 16,384 tokens for the response.

All experiments are conducted on 8 H800-80G GPUs. The supervised fine-tuning (SFT) stage requires approximately 4 hours, while the RL stage requires about 50 hours. All generated programs are executed on an Ubuntu 20.04 LTS server equipped with two 16-core Intel(R) Xeon(R) Gold 6444Y processors (45M cache, 3.60 GHz base clock) and 256 GiB of DDR5-4800 ECC RDIMM memory.

%% file: section/evaluation.tex
\section{Evaluation}

This section details the evaluation of \Tool{} and presents a comprehensive set of experimental results.
Specifically, we systematically evaluate \Tool{} across multiple dimensions, including its effectiveness across multiple PLs, the effectiveness of its training pipeline with \RewardName{}, the efficiency of Python-IR, and the impact of different RL reward designs through ablation studies.

\subsection{Experimental Setup}
We propose the following research questions (RQs):

\textbf{RQ1}: How effective is \Tool{} in improving transpilation capabilities across multiple PLs and on low-resource PL-related tasks?

\textbf{RQ2}: How effective is the training pipeline with the \RewardName{} of \Tool{} in enhancing the models' performance on Python-to-Others transpilation tasks?

\textbf{RQ3}: How efficient is Python as an intermediate representation?

\paragraph{Evaluation Metrics.}
We adopt the \textit{Pass@k} metric to evaluate the performance of different models. \textit{Pass@k} indicates that, given a transpilation task $(\Prog{src}, \Testset, \Lang{src}, \Lang{tar})$ and $k$ programs generated by an LLM, at least one program $\Prog{tar}$ among the $k$ programs is a valid result of the task; i.e., $\Prog{tar}$ is written in $\Lang{tar}$ and is behaviorally equivalent to $\Prog{src}$.  
In our benchmarks, $\Prog{src}$ is tested to be functionally correct modulo $\tau$, so \textit{Pass@k} in our experiments also indicates that one of the $k$ generated programs is functionally correct. In our experiments, we evaluate models' \textit{Pass@1} performance.

\paragraph{LLM Settings.}
In addition to the model we train, we evaluate existing mainstream LLMs, which vary in key characteristics such as affiliated institutions, the number of parameters, open-source availability, data cutoff dates, and pretraining objectives.
For all evaluated models, we set the temperature to 0 and use greedy decoding to ensure consistent evaluation. To ensure a fair comparison, we access models that we do not train through their official APIs and serve the models that we train using VLLM~\cite{vllm}. For each evaluation, we measure the \textit{Pass@1} of each model three times and report the average.

\paragraph{Dataset and benchmarks.}

We construct \Dataset{}, a file-level transpilation dataset with about 1 million Python-to-Others transpilation tasks built upon PyEdu-R~\cite{li2025codeio, benallal2024smollmcorpus, lozhkov2024starcoderstackv2}, a dataset comprising approximately 369K complex code files across categories such as algorithms, logic puzzles, math-related tasks, scientific computation, and system modeling.
The construction of \Dataset{} begins by extracting Python code snippets from PyEdu-R and standardizing their I/O into JSON format.
Subsequently, we leverage LLMs to automatically synthesize test inputs for each program using prompts that explicitly account for edge cases.
To improve execution reliability, we run each program twice and discard programs that produce inconsistent outputs across runs. While this heuristic does not guarantee determinism, it effectively filters out many non-deterministic programs, yielding a refined set of source programs, $\Prog{src}$.
These source Python programs $\Prog{src}$ are subsequently used to construct the transpilation task by designating a target PL in the target PLs, \FullTargetLangsList{}.
Through the iterative application of this pipeline, we obtain \Dataset{}, which comprises approximately 1 million Python-to-Others transpilation tasks with problem-diverse and high-quality Python source programs, which not only encompass a broad spectrum of computational domains, but also incorporate complex programming constructs, such as external API invocations and exception handling, capturing the inherent intricacies of real-world program migration scenarios.

\begin{definition}[Sampling Weight Function]
Given a program $p$, let $P(p)$ denote the specific problem solved by $p$, and let $C(\cdot)$ be a function that maps a problem to its corresponding problem class. We define the sampling weight function as:
\begin{equation*}
w(p) = \Bigl\vert \big\{ (\Prog{src}, \Testset, \Lang{src}, \Lang{tar}) \in \Datasetnm{} \mid C(P(\Prog{src})) \neq C(P(p)) \big\} \Bigr\vert.
\end{equation*}
\end{definition}

\paragraph{Sampling Strategy.} To construct the task subset, we sample \Dataset{} without replacement based on the defined weight function. We restrict our selection to transpilation tasks where $\Prog{src}$ has over 50 lines of code, distributed uniformly across the target PLs.

For the training corpus, we sample fewer than 30K tasks from \Dataset{} to form \Distilldataset{} and \RLdataset{} due to computational overhead.
To systematically evaluate our approach, we construct benchmarks for the following transpilation tasks:
\begin{itemize}
    \item Python-to-Others tasks: We use \Benchmark{} to evaluate models' performance on transpilation tasks from Python to 9 other PLs.
We construct the benchmark by sampling a subset of 900 transpilation tasks from \Dataset{}, excluding the training data. This subset comprises an equal number of tasks for each PL and features problem-diverse, structurally complex Python source code.
    \item Others-to-All tasks: We use \Benchmarkothers{} to evaluate generalized transpilation performance from non-Python PLs to an entire suite of 10 PLs, including Python.
    To construct this benchmark, we sample a subset of Python-to-Others tasks from \Dataset{} to distill \Powerfulllm{} and obtain the corresponding translated programs.
    We then filter the translated programs by execution results, yielding 100 oracle programs for each of the 9 target PLs.
These oracle programs subsequently serve as the source programs for the Others-to-All benchmark.
By pairing these source programs across the 9 PLs with 10 target PLs, which include Python, we obtain a comprehensive suite of 8.1K unique Others-to-All transpilation tasks.
\end{itemize}

These two benchmarks exemplify the multilingual transpilation domain, distinguished by their extensive problem variety and structural diversity.
They encompass 10 distinct PLs that span a broad spectrum of programming paradigms, type systems, and resource levels.
Sampled from the \Dataset{}, the constituent programs in the benchmarks extend beyond fundamental algorithmic logic to incorporate real-world programming complexities, including exception handling and external API interactions for tasks such as container management and file I/O operations, thereby reflecting practical program migration scenarios.
In contrast to existing benchmarks, which are typically limited to function- or class-level transpilation tasks~\cite{multipl-e, xue2025classeval}, cover only a narrow and homogeneous set of PLs~\cite{ahmad2023avatar, Unsupervised_translation, zheng2023codegeex}, or fail to capture sufficient problem complexity~\cite{ahmad2023avatar, Unsupervised_translation}, our proposed benchmarks provide a more problem-diverse and multilingual evaluation framework for assessing the multilingual transpilation capabilities of LLMs.

\input{tables/others2others_table}

\subsection{Effectiveness of \Tool{}}

\input{figures/resource_type_bars}

In this section, we conduct an experiment studying the effectiveness of \Tool{} in improving transpilation capabilities across multiple PLs and on low-resource PL-related tasks.

By exclusively training the \Base{} model, \Basemodel{}, on Python-to-Others transpilation tasks, we obtain an SFT variant trained on \Distilldataset{}, \BaseSft{}, and an RL variant, \RLPartial{}, which is further optimized via \RewardName{}-augmented RL on \RLdataset{}.
We subsequently evaluate the zero-shot generalization capabilities of these models on Others-to-All transpilation tasks, without any direct training, using the benchmark \Benchmarkothers{}.

\autoref{tab:others2others} presents the performance of six models: \Base{}, \BaseSft{}, \RLPartial{}, and three baseline models, namely Qwen3-235B-A22B-Instruct-2507, DeepSeek-V3.1, and Qwen3-Next-80B-A3B-Instruct.
Compared with \Base{}, \BaseSft{} consistently improves performance across all Others-to-All tasks and yields an average gain of 49.87\%.
Relative to \BaseSft{}, \RLPartial{} improves performance on 98.77\% of Others-to-All tasks, with only one task remaining unchanged.
Overall, the RL stage yields an average improvement of 9.94\%.
These results suggest that Python serves as an effective IR for enhancing code understanding across multiple PLs.

Despite possessing hundreds of billions fewer parameters, \RLPartial{} outperforms the baseline models Qwen3-235B-A22B-Instruct-2507, DeepSeek-V3.1, and Qwen3-Next-80B-A3B-Instruct on 91.36\%, 90.12\%, and 93.83\% of the evaluated transpilation tasks, respectively.
Furthermore, it achieves average performance improvements of 15.19\%, 20.83\%, and 21.22\% relative to each respective baseline.

This substantial gain underscores the effectiveness of \Tool{} in systematically enhancing a model's general transpilation capabilities.
However, the baselines retain an advantage in specific tasks involving translation into dynamically typed languages, such as \textsf{Python}, \textsf{Perl}, and \textsf{JavaScript}.
We attribute this to their structural flexibility, which renders them more akin to natural language~\cite{pandey2024towards}—a domain where massive, mainstream LLMs inherently excel.
Furthermore, \RLPartial{}'s underperformance in certain Rust transpilation tasks can be ascribed to the PL's paradigm-blending design~\cite{yun2025beyond} and complex feature set~\cite{matsakis2014rust}, which introduce intricate semantic challenges that hinder the model's learning process.

\paragraph{Performance on low-resource PL-related tasks.}
\autoref{fig:high-low-resource} presents a performance comparison across low-resource PL-related transpilation tasks.
These tasks are categorized into High-to-Low (H-L), Low-to-High (L-H), and Low-to-Low (L-L) transpilation directions, bridging high-resource PLs, such as \textsf{Python}, \textsf{C++}, \textsf{Java}, and \textsf{JavaScript}, with low-resource PLs, such as \textsf{Rust}, \textsf{Perl}, and \textsf{Haskell}~\cite{multipl-e, tiobe, cassano2024knowledge}.
The evaluation results indicate that both \BaseSft{} and \RLPartial{} consistently enhance performance on all low-resource PL-related tasks. In particular, \BaseSft{} outperforms \Base{} by an average margin of 42\%, while \RLPartial{} yields a further average increase of 7.99\% relative to \BaseSft{}.
Notably, \RLPartial{} outperforms the baseline models, Qwen3-235B-A22B-Instruct-2507, DeepSeek-V3.1, and Qwen3-Next-80B-A3B-Instruct, by considerable margins of 10.67\%, 14.09\%, and 18.16\%, respectively.
Collectively, these findings indicate the effectiveness of \Tool{} in advancing models' low-resource transpilation capabilities while mitigating the challenge of data scarcity for low-resource PLs.

\textbf{Result for RQ1:}
\Tool{} is effective in bootstrapping the model's transpilation capabilities both across multiple PLs and on low-resource PL-related tasks without requiring parallel corpora.
When trained exclusively on the Python-to-Others tasks, both \BaseSft{} and \RLPartial{} achieve average improvements of 49.87\% and 59.81\% on the Others-to-All tasks, with \RLPartial{} outperforming the baseline models Qwen3-235B-A22B-Instruct-2507, DeepSeek-V3.1, and Qwen3-Next-80B-A3B-Instruct by 15.19\%, 20.83\%, and 21.22\%, respectively.
For low-resource PL-related transpilation tasks, \BaseSft{} achieves a 42.00\% average improvement relative to \Base{}, while \RLPartial{} yields an additional 7.99\% gain and exceeds the baseline models by averages of 10.67\%, 14.09\%, and 18.16\%, respectively.

\input{figures/scatter}

\subsection{Effectiveness of \Tool{}'s training pipeline with \RewardName{}}
\input{tables/py2others_table}

In this section, we directly evaluate the effectiveness of the training pipeline of \Tool{}, including the Python-IR-pivoted SFT stage and the \RewardName{} augmented RL stage, in enhancing the models' performance on Python-to-Others transpilation tasks in \Benchmark{}.
We do not evaluate Python as an IR.

\autoref{tab:main-results} presents the performance of \RLPartial{} and mainstream LLMs on Python-to-Others transpilation tasks.
Across all evaluated models, \RLPartial{} achieves the highest overall \textit{Pass@1} on \Benchmark{}.
It outperforms other mainstream LLMs, such as DeepSeek-R1 (671B) and Claude-Sonnet-4, by 5.34\% and 14.08\% in overall \textit{Pass@1}, respectively.
This indicates that \RLPartial{} has the best multilingual transpilation capability.
In particular, \RLPartial{} delivers the best results in 6 out of 9 Python-to-Others transpilation tasks and yields an average improvement of 6.78\% over the second-best model for each PL.
We attribute these gains to the effectiveness of the training pipeline of \Tool{}, which enables a smaller model to better understand programs and to develop stronger reasoning-driven transpilation capabilities across multiple PLs.

For the Python-to-JavaScript transpilation task, \RLPartial{} exhibits a 2.67\% decrease in \textit{Pass@1} relative to the best-performing model.
We attribute this to the high-resource nature of \textsf{JavaScript}, which constitutes a large share of the training corpora of current LLMs.
\RLPartial{} does not achieve the best results for the Python-to-Golang and Python-to-Rust transpilation tasks.
This is attributed to the fact that Golang and Rust are modern multi-paradigm languages~\cite{yun2025beyond} with paradigm-blending designs that combine functional, procedural, and object-oriented approaches.
Such paradigm blending increases syntactic and grammatical complexity, particularly in Rust~\cite{matsakis2014rust}, often requiring more elaborate constructs to express equivalent semantics and introducing substantial divergence from other languages.
This hinders the easy conversion of partially correct programs into functionally correct ones.
Despite this, compared to the initial \textit{Pass@1} performance of \Base{}, the performance of \RLPartial{} improves by 1671.77\% on the Python-to-Golang task and by 1555.67\% on the Python-to-Rust task, further indicating the effectiveness of the \Tool{} training pipeline with \RewardName{}.

\autoref{fig:scatter} presents the relationship between model size and \textit{Pass@1} across open-source models.
It indicates that the model trained with the \Tool{} pipeline attains the highest \textit{Pass@1} with the fewest parameters, underscoring the pipeline's effectiveness.

\textbf{Result for RQ2:}
The training pipeline with \RewardName{} is effective. \RLPartial{} achieves the highest \textit{Pass@1} rate of 64.67\% on the Python-to-Others tasks, outperforming leading mainstream LLMs, such as DeepSeek-R1 at 59.33\%, despite having hundreds of billions fewer parameters.

\subsubsection{Ablation of Different RL Rewards}

\input{tables/ablation_table}

We then conduct an ablation study to evaluate the effectiveness of the \RewardName{} in RL training in comparison to alternative rewards using the same training corpus, \RLdataset{}.
Building on \BaseSft{}, we use the following rewards to train three RL variants:

\begin{itemize}
    \item Linear execution reward: This reward increases linearly with the number of passed test cases. Formally, we use the reward function in \autoref{eq:reward} and set the reward gate function to $\sigma(x)=x$.
    \item Conservative-Partial-Functional reward: In contrast to the \emph{Aggressive-Partial-Functional} reward, this reward reduces the extent to which we reward those rollouts $o \sim \pi_\theta(\cdot \mid q)$ such that $o$ contains a program $\Prog{tar}$ in the target language and $\Prog{tar}$ is partially functionally correct.
    Therefore, we emphasize rollouts $o' \sim \pi_\theta(\cdot \mid q)$ that contain a functionally correct target program. Thus, when we use the reward function in \autoref{eq:reward}, the reward gate function is a convex function defined as:
    \begin{align}
\sigma(x) = \frac{\lambda^{x}-1}{\lambda-1}, \text{ where } \lambda > 1,
\end{align}
where $\lambda$ is a hyperparameter that controls the extent to which we assign rewards for partial functional correctness.
    \item Discrete reward: This reward is an extreme variant of the Conservative-Partial-Functional reward. It assigns reward only to rollouts $o \sim \pi_\theta(\cdot \mid q)$ such that $o$ contains a program $\Prog{tar}$ in the target language and $\Prog{tar}$ is functionally correct. Thus, it assigns zero reward to rollouts $o' \sim \pi_\theta(\cdot \mid q)$ that contain a functionally incorrect or partially functionally correct target program. Formally, we use the reward function in \autoref{eq:reward} and define the reward gate function as: $\sigma(x)=[x=1]$.
\end{itemize}

\autoref{tab:ablation-results} presents the performance of the SFT baseline \BaseSft{} and four RL variants trained on top of it with distinct reward functions.
All four RL variants surpass the \BaseSft{} baseline on every Python-to-Others transpilation task, indicating the effectiveness of RL for improving code transpilation performance.
Among the variants, \RLPartial{} achieves the best overall performance with the highest average \textit{Pass@1} of 64.67\%, and it delivers the top results in 7 of 9 Python-to-Others tasks, with an average improvement of 3.48\% over the second-best model for each language.
These findings indicate that \RewardName{} yields consistent gains on most Python-to-Others transpilation tasks.
In terms of overall \textit{Pass@1} improvement from RL, \RewardName{} yields the highest gain of 12.27\%.
This corresponds to absolute gains of 3.41\% over the linear execution reward, 3.78\% over the Conservative-Partial-Functional reward, and 4.11\% over the discrete reward, corresponding to 36.67\%, 42.33\%, and 47.79\% larger RL-stage relative improvements, respectively.
These results further indicate the effectiveness of \RewardName{}.

The second-best model, \RLLinear{}, achieves an overall \textit{Pass@1} that is lower than that of \RLPartial{}, yet it leads in Python-to-Rust tasks.
We attribute this to the complexity of Rust’s syntax~\cite{yun2025beyond, matsakis2014rust}, which makes it difficult to convert rollouts with partially functionally correct programs into those with functionally correct ones.
Although more rollouts contain partially functionally correct programs, the number of rollouts with functionally correct ones does not increase.
The linear reward provides the best balance between converting rollouts with functionally incorrect programs into those with partially functionally correct ones and further converting them into rollouts with functionally correct programs.
\RLFlip{} attains the best result only on Python-to-Perl with a \textit{Pass@1} of 74.00\% and reaches an overall \textit{Pass@1} of 59.80\%.
We attribute this to Perl’s relatively simple syntax and grammar and to its high initial performance (64.00\%), which facilitates the generation of many rollouts with partially functionally correct programs.
These rollouts are easy to convert into those with functionally correct programs.

Despite this, \RewardName{} delivers the largest improvement on the Python-to-Haskell task, achieving a \textit{Pass@1} rate of 35.33\%, which is 8\% higher than that achieved with the linear execution reward and corresponds to a gain that is 200\% larger than that obtained with the linear reward.
We attribute this result to the relatively low baseline performance of Base-SFT, which attains only a 23.33\% \textit{Pass@1} rate and is the lowest among all tasks, leaving substantial room for generating rollouts that contain partially correct target programs that receive positive feedback. \RewardName{} further amplifies this effect by emphasizing partial functional correctness, thereby encouraging broader exploration of such rollouts.
These results further indicate that \RewardName{} is particularly effective in improving performance on tasks that start with weaker initial performance.

\textbf{Results of the Ablation Study:}
\RewardName{} is effective in improving models' transpilation abilities, with \RLPartial{} attaining the highest \textit{Pass@1} in 7 of the 9 target languages for Python-to-Others tasks and achieving the best overall \textit{Pass@1}. Compared to \RLLinear{}, \RLPartial{} delivers a 3.41\% absolute performance gain and a 36.67\% relative improvement during the RL stage.

\subsection{Efficiency of Python as an Intermediate Representation}

\input{tables/new_py2others_table}

We conduct an experiment evaluating the efficiency of using Python as an IR to enhance models’ general transpilation capabilities.
We perform SFT on the base model \Basemodel{} using training data that contains either Python-to-Others or Any-to-Any transpilation pairs, and we compare the resulting models’ \textit{Pass@1} performance on the Others-to-All benchmark \Benchmarkothers{}.

Based on the distilled SFT dataset \Distilldataset{}, which contains approximately 20K Python-to-Others pairs, we directly construct the Python-to-Others SFT dataset using the full set.
For each Python source program in \Distilldataset{}, we group all functionally correct target programs and synthesize Any-to-Any pairs by permuting programs across languages within each group, which yields approximately 160K Any-to-Any pairs for the Any-to-Any SFT dataset.
These samples are not guaranteed to be balanced across language pairs because when distilling \Powerfulllm{} to construct \Distilldataset{}, we may not obtain functionally correct target programs for every target language for each source program.

We fine-tune the base model on each dataset for three epochs, yielding two models.
\autotabref{tab:data-type} presents the \textit{Pass@1} performance of these models on \Benchmarkothers{}.
The model fine-tuned on 20K Python-to-Others pairs outperforms the model trained on 160K Any-to-Any pairs by 5.83\%.
This result indicates the efficiency of the Python IR in enhancing the models' general transpilation capabilities.

\textbf{Result for RQ3:}
Python-IR is efficient in enhancing models’ general transpilation capabilities. Specifically, fine-tuning the base model exclusively on 20K Python-to-Others pairs yields a model that outperforms one trained on 160K Any-to-Any pairs by 5.83\% on Others-to-All tasks.

%% file: tables/others2others_table.tex
\begin{table*}[t]
\centering
\caption{
Performance comparison in \textit{Pass@1} (\%) of models on \Benchmarkothers{}, where each task translates a program from a source PL listed in the columns from \textbf{C++} to \textbf{Haskell} into a target PL listed in the first column, \textbf{Languages}.
\Base{} denotes the base model, and \BaseSft{} and \RLPartial{} denote the resulting models further trained.
Values highlighted in bold indicate the highest \textit{Pass@1} rate for each transpilation task.
}
\label{tab:others2others}
\resizebox{0.95\linewidth}{!}{
\begin{tabular}{c|l|rrrrrrrrr}
\toprule
\makecell[l]{\textbf{Languages}} & \LangGroup{Model} & \makecell[c]{\textbf{From}\\ \textbf{C++}} & \makecell[c]{\textbf{From}\\ \textbf{Java}} & \makecell[c]{\textbf{From}\\ \textbf{C\#}} & \makecell[c]{\textbf{From}\\ \textbf{JS}} & \makecell[c]{\textbf{From}\\ \textbf{Golang}} & \makecell[c]{\textbf{From}\\ \textbf{Rust}} & \makecell[c]{\textbf{From}\\ \textbf{Perl}} & \makecell[c]{\textbf{From}\\ \textbf{Ruby}} & \makecell[c]{\textbf{From}\\ \textbf{Haskell}} \\
\midrule
\multirow{7}{*}{\makecell[c]{\textbf{To}\\ \textbf{Python}}}
& Qwen3-235B-A22B-Instruct-2507 & 90.33 & 92.67 & 96.00 & 94.67 & 91.00 & 92.33 & 95.67 & 90.00 & \textbf{92.67} \\
& DeepSeek-V3.1 & 93.00 & \textbf{97.67} & 95.33 & 95.00 & 92.00 & 96.00 & 94.67 & \textbf{98.33} & 92.33 \\
& Qwen3-Next-80B-A3B-Instruct & 93.33 & 94.00 & 95.00 & 93.00 & 90.33 & \textbf{97.00} & 78.67 & 96.00 & 88.00 \\
& \Base{} (\Basemodel{}) & 65.00 & 54.67 & 71.67 & 66.00 & 62.67 & 57.33 & 68.33 & 71.00 & 46.00 \\
& \BaseSft{} & 89.33 & 86.67 & 84.00 & 82.33 & 83.00 & 72.33 & 88.33 & 88.00 & 77.00 \\
& \textbf{\RLPartial{}} & \textbf{93.67} & 95.00 & \textbf{96.33} & \textbf{95.33} & \textbf{92.33} & 83.67 & \textbf{96.33} & 90.67 & 83.33 \\
\midrule
\multirow{7}{*}{\makecell[l]{\textbf{To}\\ \textbf{C++}}}
& Qwen3-235B-A22B-Instruct-2507 & N/A & 57.00 & 63.33 & 51.67 & 55.67 & 52.33 & 45.00 & 62.67 & 56.33 \\
& DeepSeek-V3.1 & N/A & 46.67 & 16.00 & 17.67 & 56.67 & 30.00 & 10.00 & 8.33 & 36.33 \\
& Qwen3-Next-80B-A3B-Instruct & N/A & 43.33 & 50.00 & 54.00 & 53.33 & 60.00 & 49.33 & 43.67 & 52.33 \\
& \Base{} (\Basemodel{}) & N/A & 0.00 & 0.00 & 0.00 & 0.00 & 0.00 & 0.67 & 0.00 & 0.00 \\
& \BaseSft{} & N/A & 62.00 & 68.67 & 76.33 & 78.33 & 73.00 & 74.00 & 67.33 & 64.00 \\
& \textbf{\RLPartial{}} & N/A & \textbf{76.67} & \textbf{87.67} & \textbf{80.33} & \textbf{82.33} & \textbf{87.00} & \textbf{76.67} & \textbf{69.00} & \textbf{75.00} \\
\midrule
\multirow{7}{*}{\makecell[c]{\textbf{To}\\ \textbf{Java}}}
& Qwen3-235B-A22B-Instruct-2507 & 60.67 & N/A & 45.33 & 40.67 & 54.67 & 50.00 & 58.33 & 41.00 & 56.33 \\
& DeepSeek-V3.1 & 46.00 & N/A & 45.67 & 37.00 & 31.67 & 38.67 & 39.67 & 42.00 & 43.00 \\
& Qwen3-Next-80B-A3B-Instruct & 51.67 & N/A & 42.00 & 41.67 & 36.00 & 59.00 & 45.00 & 42.00 & 42.00 \\
& \Base{} & 30.33 & N/A & 15.00 & 9.67 & 24.00 & 20.00 & 10.00 & 8.00 & 19.00 \\
& \BaseSft{} & 66.67 & N/A & 39.00 & 77.00 & 81.00 & 76.33 & 70.67 & 71.33 & 63.33 \\
& \textbf{\RLPartial{}} & \textbf{82.33} & N/A & \textbf{66.67} & \textbf{77.00} & \textbf{87.00} & \textbf{84.33} & \textbf{75.67} & \textbf{74.67} & \textbf{73.33} \\
\midrule
\multirow{7}{*}{\makecell[c]{\textbf{To}\\ \textbf{C\#}}}
& Qwen3-235B-A22B-Instruct-2507 & 33.67 & 28.00 & N/A & 54.33 & 45.67 & 63.00 & 63.67 & 50.33 & 54.33 \\
& DeepSeek-V3.1 & 49.67 & 57.67 & N/A & 54.33 & 35.33 & 52.00 & 54.33 & 56.33 & 56.33 \\
& Qwen3-Next-80B-A3B-Instruct & 49.00 & 22.00 & N/A & 46.33 & 60.67 & 54.33 & 45.33 & 27.00 & 50.33 \\
& \Base{} (\Basemodel{}) & 1.33 & 0.67 & N/A & 1.67 & 0.00 & 0.33 & 1.67 & 1.33 & 0.33 \\
& \BaseSft{} & 55.00 & 62.67 & N/A & 71.33 & 51.33 & 66.67 & 65.33 & 61.67 & 52.67 \\
& \textbf{\RLPartial{}} & \textbf{66.00} & \textbf{72.00} & N/A & \textbf{78.33} & \textbf{75.67} & \textbf{69.67} & \textbf{72.33} & \textbf{66.33} & \textbf{67.33} \\
\midrule
\multirow{7}{*}{\makecell[c]{\textbf{To}\\ \textbf{JS}}}
& Qwen3-235B-A22B-Instruct-2507 & 91.67 & 81.33 & 90.00 & N/A & 81.67 & 90.33 & \textbf{87.33} & 76.67 & \textbf{84.67} \\
& DeepSeek-V3.1 & 87.33 & 73.00 & 71.00 & N/A & 84.00 & 78.67 & 70.67 & 54.00 & 83.33 \\
& Qwen3-Next-80B-A3B-Instruct & 89.00 & 82.33 & 91.00 & N/A & 81.00 & 88.67 & 85.67 & 79.67 & 83.33 \\
& \Base{} (\Basemodel{}) & 29.67 & 15.67 & 18.00 & N/A & 31.33 & 21.33 & 10.33 & 7.33 & 11.33 \\
& \BaseSft{} & 84.00 & 77.33 & 83.67 & N/A & 77.33 & 83.67 & 80.67 & 78.00 & 74.67 \\
& \textbf{\RLPartial{}} & \textbf{93.00} & \textbf{82.33} & \textbf{91.33} & N/A & \textbf{85.00} & \textbf{92.00} & 83.67 & \textbf{81.33} & 78.67 \\
\midrule
\multirow{7}{*}{\makecell[c]{\textbf{To}\\ \textbf{Golang}}}
& Qwen3-235B-A22B-Instruct-2507 & 37.67 & 38.67 & 30.67 & 27.33 & N/A & 40.67 & 29.33 & 25.00 & 37.33 \\
& DeepSeek-V3.1 & 28.33 & 35.33 & 36.33 & 38.00 & N/A & 27.00 & 24.00 & 33.33 & 28.67 \\
& Qwen3-Next-80B-A3B-Instruct & 27.00 & 27.67 & 30.00 & 24.00 & N/A & 27.67 & 19.00 & 25.33 & 26.33 \\
& \Base{} (\Basemodel{}) & 6.00 & 13.33 & 4.33 & 7.33 & N/A & 21.67 & 6.00 & 8.33 & 9.00 \\
& \BaseSft{} & 58.00 & 56.67 & 65.00 & 67.33 & N/A & 62.00 & 61.67 & 62.33 & 59.00 \\
& \textbf{\RLPartial{}} & \textbf{64.00} & \textbf{61.67} & \textbf{75.00} & \textbf{68.00} & N/A & \textbf{71.33} & \textbf{74.00} & \textbf{65.67} & \textbf{60.33} \\
\midrule
\multirow{7}{*}{\makecell[c]{\textbf{To}\\ \textbf{Rust}}}
& Qwen3-235B-A22B-Instruct-2507 & 57.33 & 57.00 & 59.33 & 53.33 & 60.67 & N/A & 64.67 & \textbf{60.67} & \textbf{70.33} \\
& DeepSeek-V3.1 & \textbf{66.67} & 60.00 & 43.33 & 37.67 & 61.00 & N/A & 64.33 & 57.00 & 42.00 \\
& Qwen3-Next-80B-A3B-Instruct & 54.67 & 38.33 & 46.33 & 37.67 & 49.00 & N/A & 39.00 & 44.67 & 48.00 \\
& \Base{} (\Basemodel{}) & 9.00 & 9.33 & 15.33 & 9.33 & 14.00 & N/A & 8.67 & 12.00 & 14.67 \\
& \BaseSft{} & 59.67 & 54.67 & 56.00 & 52.67 & 45.67 & N/A & 50.33 & 55.33 & 53.33 \\
& \textbf{\RLPartial{}} & 61.00 & \textbf{60.33} & \textbf{65.67} & \textbf{58.33} & \textbf{61.33} & N/A & \textbf{65.00} & 57.33 & 62.00 \\
\midrule
\multirow{7}{*}{\makecell[c]{\textbf{To}\\ \textbf{Perl}}}
& Qwen3-235B-A22B-Instruct-2507 & 66.00 & 65.67 & 64.67 & 64.00 & 60.00 & \textbf{80.33} & N/A & 64.67 & 68.33 \\
& DeepSeek-V3.1 & \textbf{82.00} & 69.33 & 81.67 & 80.67 & 75.67 & 78.67 & N/A & 71.33 & 70.67 \\
& Qwen3-Next-80B-A3B-Instruct & 72.33 & 63.67 & 73.67 & 67.00 & 69.33 & 68.00 & N/A & 58.67 & 69.33 \\
& \Base{} (\Basemodel{}) & 29.33 & 27.67 & 40.00 & 32.00 & 34.00 & 39.00 & N/A & 19.67 & 11.67 \\
& \BaseSft{} & 73.67 & 64.33 & 72.00 & 75.33 & 65.33 & 64.00 & N/A & 66.00 & 54.00 \\
& \textbf{\RLPartial{}} & 74.00 & \textbf{70.33} & \textbf{83.33} & \textbf{81.00} & \textbf{76.33} & 69.00 & N/A & \textbf{72.00} & \textbf{71.00} \\
\midrule
\multirow{7}{*}{\makecell[c]{\textbf{To}\\ \textbf{Ruby}}}
& Qwen3-235B-A22B-Instruct-2507 & 66.67 & 65.67 & 69.33 & 58.00 & 55.33 & 66.67 & 59.00 & N/A & 58.67 \\
& DeepSeek-V3.1 & 44.00 & 57.33 & 62.33 & 25.67 & 46.00 & 45.00 & 43.33 & N/A & 62.00 \\
& Qwen3-Next-80B-A3B-Instruct & 57.00 & 55.00 & 58.67 & 49.00 & 53.00 & 51.67 & 43.00 & N/A & 58.00 \\
& \Base{} (\Basemodel{}) & 51.00 & 47.33 & 40.67 & 43.67 & 42.00 & 43.00 & 41.67 & N/A & 25.67 \\
& \BaseSft{} & 78.00 & 72.67 & 60.00 & 74.67 & 68.00 & 63.00 & 71.00 & N/A & 54.67 \\
& \textbf{\RLPartial{}} & \textbf{84.33} & \textbf{83.33} & \textbf{78.00} & \textbf{86.00} & \textbf{78.00} & \textbf{78.00} & \textbf{80.00} & N/A & \textbf{67.67} \\
\midrule
\multirow{7}{*}{\makecell[c]{\textbf{To}\\ \textbf{Haskell}}}
& Qwen3-235B-A22B-Instruct-2507 & 10.33 & 15.67 & 17.00 & 9.67 & 12.67 & 20.33 & 11.00 & 11.33 & N/A \\
& DeepSeek-V3.1 & 15.67 & 12.00 & 12.33 & 10.00 & 8.67 & 17.33 & 14.33 & 10.33 & N/A \\
& Qwen3-Next-80B-A3B-Instruct & 0.67 & 1.00 & 1.00 & 0.00 & 1.00 & 2.67 & 1.00 & 0.00 & N/A \\
& \Base{} (\Basemodel{}) & 0.00 & 0.00 & 0.00 & 0.00 & 0.00 & 0.00 & 0.00 & 0.00 & N/A \\
& \BaseSft{} & 30.33 & 25.00 & 21.00 & 29.00 & 22.33 & 36.33 & 21.00 & 24.67 & N/A \\
& \textbf{\RLPartial{}} & \textbf{31.33} & \textbf{28.67} & \textbf{36.00} & \textbf{41.00} & \textbf{37.67} & \textbf{43.33} & \textbf{31.33} & \textbf{34.33} & N/A \\
\bottomrule
\end{tabular}
}
\end{table*}

%% file: figures/resource_type_bars.tex
\begin{figure}[t]
    \centering
    \includegraphics[width=0.9\linewidth]{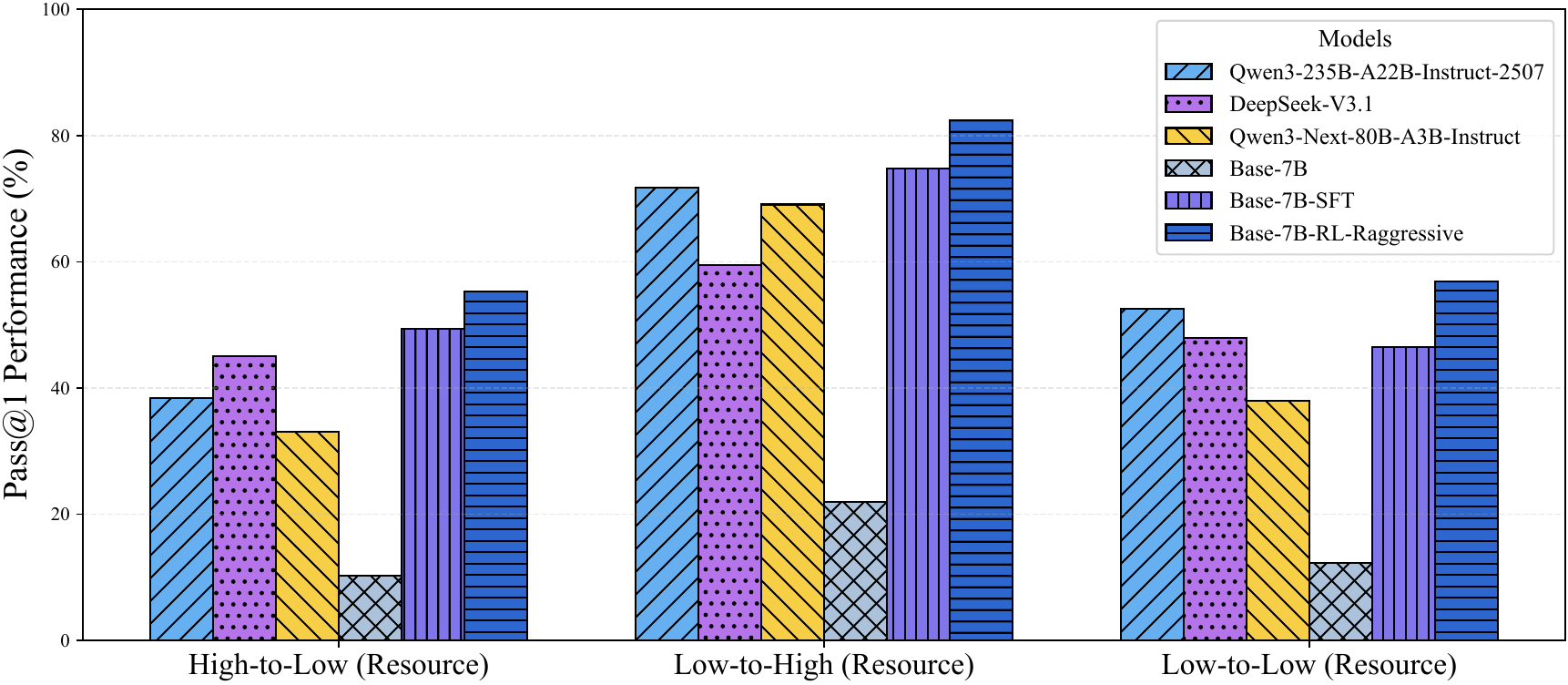}
    \captionof{figure}{Performance of models on low-resource PL-related transpilation tasks.}
    \label{fig:high-low-resource}
\end{figure}

%% file: figures/scatter.tex
\begin{wrapfigure}{l}{0.6\textwidth}
    \centering
    \includegraphics[width=\linewidth]{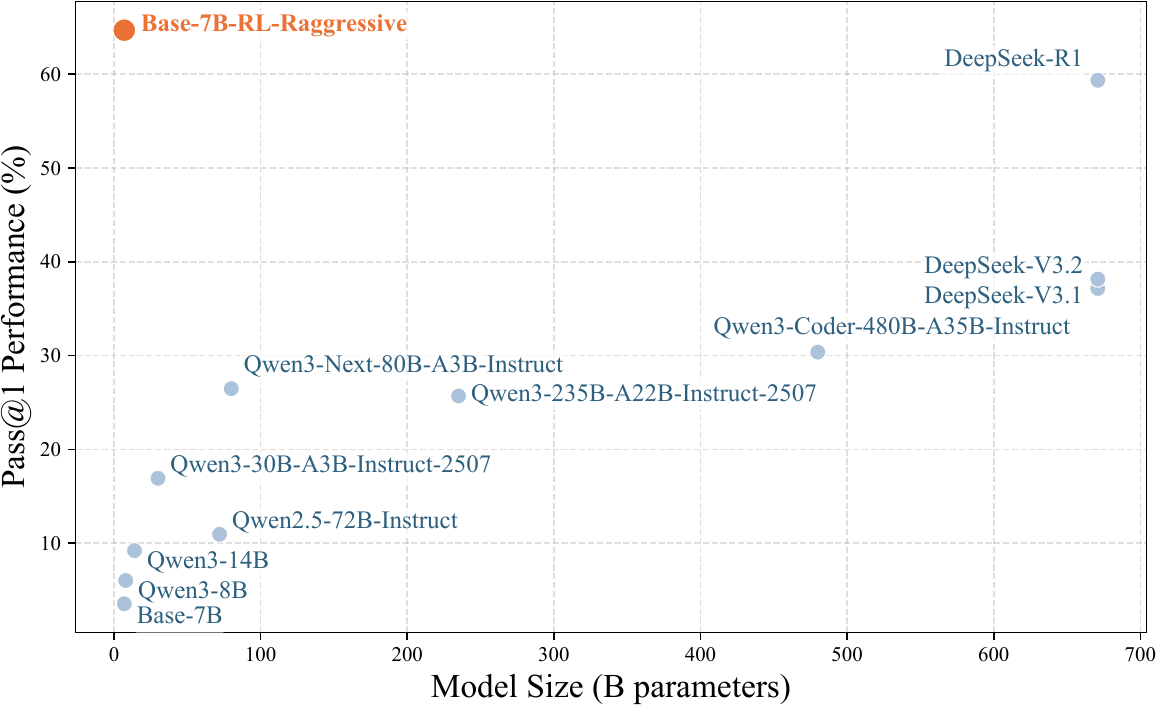}
    \caption{Scatter plot illustrating the relationship between model size (B) and \textit{Pass@1} performance (\%) on \Benchmark{}.}
    \label{fig:scatter}
\end{wrapfigure}

%% file: tables/py2others_table.tex
\begin{table*}[t]
    \centering
    \caption{
    Performance comparison in terms of \textit{Pass@1} (\%) of different models on \Benchmark{}, in which each task involves translating a program from \textbf{Python} to one of the target programming languages listed in the columns, from \textbf{C++} to \textbf{Haskell}. The \textbf{Total} column presents the overall \textit{Pass@1} score across \Benchmark{}. \RLPartial{} is the model trained using our training pipeline with the \RewardName{} from the base model, \Base{}. Bold values indicate the highest \textit{Pass@1} rate in each column. * denotes closed-source models.
}
\label{tab:main-results}
    \resizebox{\linewidth}{!}{
    \begin{tabular}{l|ccccccccc|c}
        \toprule
        \textbf{Model} & \makecell[c]{\textbf{C++}} & \makecell[c]{\textbf{Java}} & \makecell[c]{\textbf{C\#}} & \makecell[c]{\textbf{JavaScript}} & \makecell[c]{\textbf{Golang}} & \makecell[c]{\textbf{Rust}} & \makecell[c]{\textbf{Perl}} & \makecell[c]{\textbf{Ruby}} & \textbf{Haskell} & \textbf{Total} \\ 
        \midrule
        GPT-4o$^{*}$ & 17.33 & 37.00 & 27.33 & 41.33 & 33.00 & 29.33 & 45.00 & 38.00 & 0.33 & 29.85 \\ 
        GPT-4.1$^{*}$ & 24.33 & 65.00 & 60.00 & 73.33 & 51.67 & 58.67 & 68.33 & 39.00 & 24.67 & 51.67 \\ 
        Gemini-2.5-Flash$^{*}$ & 54.00 & 50.33 & 53.67 & 75.00 & 44.33 & 49.67 & 56.00 & 54.67 & 17.00 & 50.52 \\ 
        Claude-Sonnet-4$^{*}$ & 20.33 & 64.67 & 60.67 & 68.00 & 49.67 & \textbf{65.33} & 67.33 & 24.33 & 35.00 & 50.59 \\ 
        GPT-4.1-mini$^{*}$ & 15.33 & 50.33 & 44.67 & 65.67 & 46.67 & 59.33 & 66.33 & 22.67 & 30.33 & 44.59 \\ 
        \midrule 
        DeepSeek-R1 & 35.67 & 70.00 & 63.33 & \textbf{77.00} &  \textbf{67.67} & 63.00 & 67.33 & 67.67 & 22.33 & 59.33 \\ 
        Qwen3-235B-A22B-Instruct-2507 & 32.33 & 40.00 & 21.67 & 30.33 & 25.33 & 28.33 & 26.67 & 20.00 & 6.67 & 25.70 \\ 
        Qwen3-Coder-480B-A35B-Instruct & 22.33 & 41.33 & 32.67 & 43.00 & 28.67 & 33.67 & 41.33 & 19.00 & 11.33 & 30.37 \\ 
        DeepSeek-V3.1 & 13.67 & 43.33 & 42.67 & 46.00 & 34.33 & 46.00 & 48.33 & 47.67 & 12.33 & 37.15 \\ 
        DeepSeek-V3.2 & 17.33 & 47.00 & 43.67 & 49.67 & 34.33 & 45.00 & 55.00 & 43.33 & 8.00 & 38.15 \\ 
        Qwen3-Next-80B-A3B-Instruct & 25.33 & 28.00 & 23.00 & 47.67 & 13.00 & 28.67 & 33.00 & 39.33 & 0.33 & 26.48 \\ 
        Qwen2.5-72B-Instruct & 6.67 & 7.33 & 14.67 & 12.00 & 3.33 & 9.67 & 18.33 & 26.67 & 0.00 & 10.96 \\ 
        Qwen3-30B-A3B-Instruct-2507 & 10.00 & 24.33 & 13.33 & 25.00 & 7.00 & 20.33 & 23.33 & 29.00 & 0.00 & 16.93 \\ 
        Qwen3-14B & 2.33 & 5.67 & 11.67 & 10.67 & 3.67 & 4.67 & 16.00 & 28.00 & 0.33 & 9.22 \\ 
        Qwen3-8B & 2.00 & 6.67 & 7.33 & 8.33 & 3.00 & 5.00 & 14.00 & 8.00 & 0.00 & 6.04 \\ 
        \midrule
        \Base{} (\Basemodel{}) & 1.33 & 3.67 & 0.00 & 5.67 & 3.33 & 3.00 & 4.33 & 10.67 & 0.00 & 3.56 \\ 
        \RLPartial{}(Ours) & \textbf{68.67} & \textbf{70.00} & \textbf{69.00} & 74.33 & 59.00 & 49.67 & \textbf{72.00} & \textbf{84.00} & \textbf{35.33} & \cellcolor{green!15} \textbf{64.67} \\ 
        \bottomrule
    \end{tabular}
    }
\end{table*}

%% file: tables/ablation_table.tex
\begin{table*}[t]
    \centering
    \caption{Performance comparison in \textit{Pass@1} (\%) of models trained with different rewards on \Benchmark{}, where each task translates a program from \textbf{Python} to one of the target programming languages listed in the columns from \textbf{C++} to \textbf{Haskell}. The \textbf{Total} column reports the overall \textit{Pass@1} across \Benchmark{}.
    \RLLinear{} denotes the resulting model further trained from \BaseSft{} with RL on \RLdataset{} using the linear execution reward. \RLFlip{} denotes the resulting model further trained from \BaseSft{} with RL using the Conservative-Partial-Functional reward. \RLDiscrete{} denotes the resulting model further trained from \BaseSft{} with RL using the discrete reward.
Bold values indicate the highest \textit{Pass@1} rate in each column.
}
\label{tab:ablation-results}
    \resizebox{\linewidth}{!}{
    \begin{tabular}{l|ccccccccc|c}
        \toprule
        \textbf{Model} &  \makecell[c]{\textbf{C++}} & \makecell[c]{\textbf{Java}} & \makecell[c]{\textbf{C\#}} & \makecell[c]{\textbf{JavaScript}} & \makecell[c]{\textbf{Golang}} & \makecell[c]{\textbf{Rust}} & \makecell[c]{\textbf{Perl}} & \makecell[c]{\textbf{Ruby}} & \textbf{Haskell} & \textbf{Total} \\ 
        \midrule
        \BaseSft{} & 56.67 & 54.67 & 54.67 & 65.33 & 43.00 & 40.33 & 64.00 & 65.67 & 23.33 & 51.96 \\ 
        \RLLinear{} & 62.67 & 67.00 & 66.33 & 66.00 & 58.33 & \textbf{53.67} & 69.33 & 80.67 & 27.33 & 61.26 (+9.30) \\ 
        \RLFlip{} & 61.33 & 65.67 & 67.67 & 69.33 & 56.00 & 47.67 & \textbf{74.00} & 78.33 & 28.00 & 60.89 (+8.93) \\ 
        \RLDiscrete{} & 65.00 & 66.33 & 61.67 & 69.00 & 57.00 & 51.00 & 70.33 & 76.67 & 28.00 & 60.56 (+8.60) \\ 
        \RLPartial{} (ours) & \textbf{68.67} & \textbf{70.00} & \textbf{69.00} & \textbf{74.33} & \textbf{59.00} & 49.67 & 72.00 & \textbf{84.00} & \textbf{35.33} & \cellcolor{green!15} \textbf{64.67 (+12.71)} \\ 
        \bottomrule
    \end{tabular}
    }
\end{table*}

%% file: tables/new_py2others_table.tex
\begin{table*}[t]
    \centering
    \caption{
    Performance comparison in \textit{Pass@1} (\%) on \Benchmarkothers{} between two models fine-tuned on Python-to-Others tasks and Any-to-Any tasks, respectively. The models are trained using two types of data.
}
    \label{tab:new-py2others}
    \resizebox{0.5\linewidth}{!}{
      \begin{tabular}{l|c}
        \toprule
        \textbf{SFT Data Type} & \textbf{Model Performance (\%)} \\
        \midrule
        Python-to-Others pairs    & \textbf{53.85} \\
        Any-to-Any pairs & 48.02 \\
        \bottomrule
      \end{tabular}
    }
    \label{tab:data-type}
\end{table*}

%% file: section/discussion.tex
\section{Discussion}
\paragraph{Future work.}
Potential future work includes systematically investigating the application of \RewardName{} in RL training for coding-related tasks such as code generation.
In addition, the use of the high-resource PL Python as an intermediate representation to address low-resource PL transpilation warrants further exploration, particularly by leveraging such IRs to enhance the code generation capabilities of current LLMs.
Furthermore, while investigating alternative PLs as potential IRs remains a compelling direction, the dearth of comparable high-quality corpora for other PLs constrains current efforts, and we therefore leave this exploration to future work.
Additionally, while \Dataset{} provides high-quality, complex file-level transpilation tasks, it intentionally excludes repository-level complexities, such as multi-file structures and build systems, which are beyond the current scope and are left for future work.

\paragraph{Threats to validity.}
In our evaluation and training, we rely on currently popular mainstream large language models, which may limit the validity of our conclusions for future model generations.
To mitigate this risk, we evaluate multiple models from different institutions with varying parameter scales and open-source availability, ensuring that our findings reflect a broad set of models.
We train the model on only 10 PLs that are mainstream at present, and it remains unclear whether the approach can scale to a substantially larger set of languages.
To mitigate this limitation, we select languages to cover diverse type systems, resource availability, and programming paradigms, and we focus on widely used and well-known languages that are broadly representative.

%% file: section/conclusion.tex
\section{Conclusion}


We present \Tool{}, a training framework based on Python as an intermediate representation (IR) that bootstraps the models' multilingual transpilation capabilities without requiring parallel corpora.
It combines Python-IR-pivoted SFT and \RewardName{} augmented RL with language-agnostic test cases.
Comprehensive experiments show the effectiveness of \Tool{} in enhancing models' transpilation capabilities.





%% file: section/appendix.tex
\appendix

\section{Training Dynamics}

\input{figures/rl_rewards}

\autoref{fig:rl_rewards} presents the trajectories of the mean reward during reinforcement learning (RL) training under four distinct reward formulations. Notably, all four curves show upward trends, indicating that the policy progressively improves code quality and test-case pass rates across all formulations.

The thresholded reward exhibits highly unstable behavior and is prone to plateaus, as shown in the highlighted region. Among the four formulations, APF achieves the highest reward throughout most of the training phase, followed by Linear Execution (LE), Conservative-Partly Functional (CPF), and Discrete rewards. This suggests that reward formulations providing denser and more continuous signals support smoother optimization and more stable training progress in program translation.
By contrast, the linear execution reward exhibits a highly unstable trajectory with a pronounced drop during the early stages, resulting in lower overall reward growth. A similar pattern is observed for CPF, whose trajectory consistently remains below that of APF throughout the training process. Both reward designs yield substantially lower value trajectories and weaker late-stage improvement.

Furthermore, the Discrete reward consistently exhibits the lowest values with a relatively flat learning curve. This stark contrast highlights the inherent challenges of sparse reward optimization. Because the Discrete strategy assigns rewards only to functionally correct outputs ($x = 1$), the resulting reward scores lack meaningful gradation during the early exploration stage. Consequently, the model finds it much more difficult to escape initially suboptimal behaviors than under formulations with richer and more informative reward signals.

\section{Prompt Template}

\input{figures/prompt-template}

%% file: figures/rl_rewards.tex
\begin{figure}[h]
    \centering
    \includegraphics[width=0.8\linewidth]{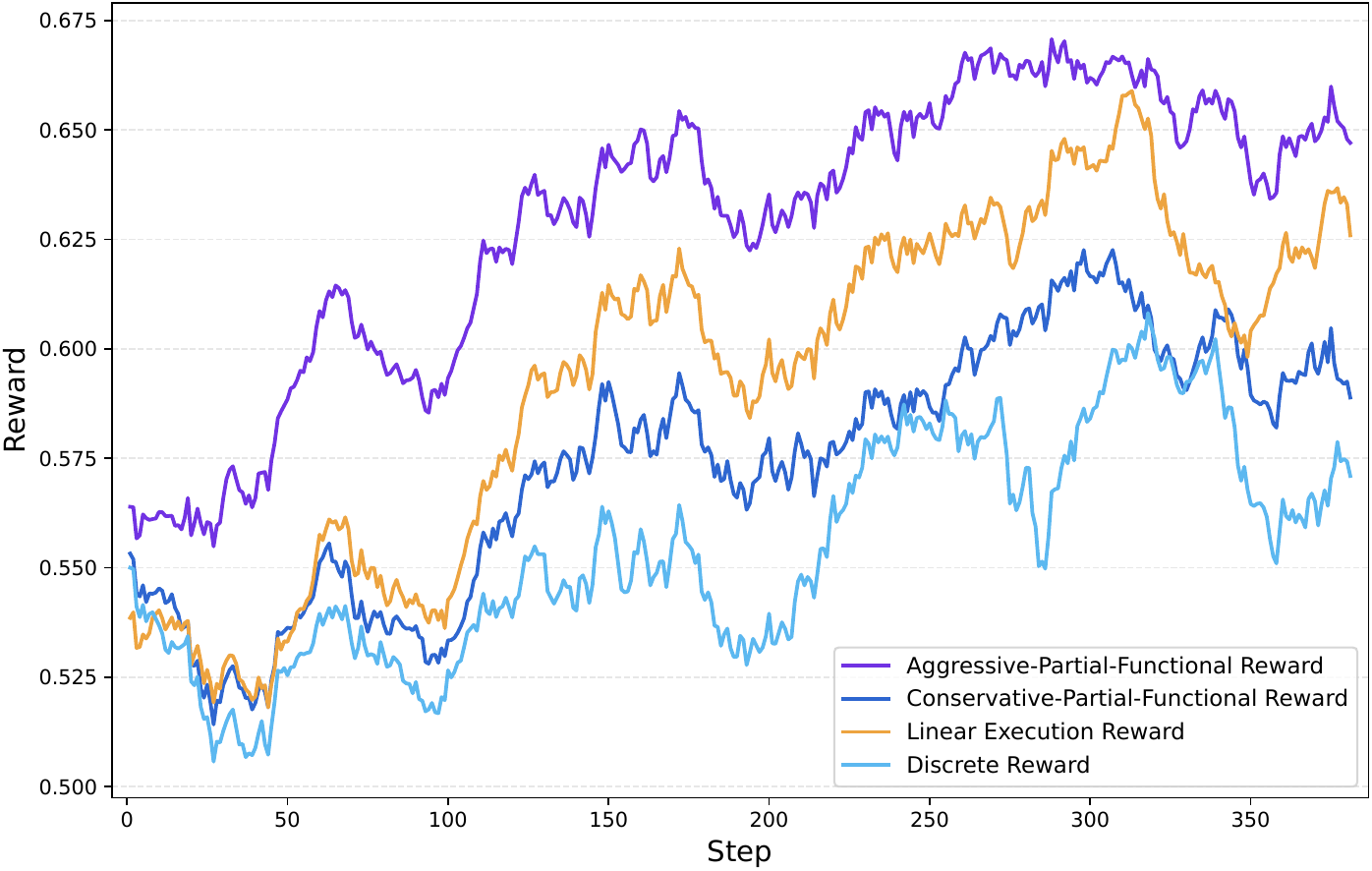}
    \caption{Comparison of mean reward trajectories during RL training under four distinct reward formulations. All curves are smoothed using an exponential moving average for clarity.
}
    \label{fig:rl_rewards}
\end{figure}

%% file: figures/prompt-template.tex
\begin{figure}[h]
    \centering
\begin{tcolorbox}[colback=white,colframe=black,boxrule=0.8pt,arc=2pt,left=6pt,right=6pt,top=6pt,bottom=6pt]
[System] You are a specialist in transpiling code between programming languages, tasked with translating the provided reference code from the source language into a functionally equivalent implementation in the target language. Ensure that variable names are preserved wherever the variables in both implementations share the same semantic meaning. If the transpiled program contains an entry class, it should be named `Main` whenever permissible.
The specifications for the code execution runtime environments and versions are outlined below. Please ensure strict adherence to these requirements:

- The Python execution environment is Python 3.8 with the standard library and the following third-party packages installed: NumPy 1.24.4, and pandas 2.0.3. For JSON parsing, use the built-in json module.

- The C\# execution environment is .NET 9 SDK (9.0.203). For JSON parsing, use the built-in standard library.

- The C++ environment is Ubuntu 20.04 with GCC 9.4.0 (g++), libstdc++ with C++17 support, and glibc 2.31 for runtime support. For JSON parsing, use the JsonCpp library and include the header <json/json.h>.

- The Java execution environment is OpenJDK 17.0.15. For JSON parsing, use the Jackson (FasterXML) library.

- The JavaScript execution environment is JavaScript Node.js v22.18.0. For JSON parsing, use the built-in JSON global object.

- The Golang execution environment is Go version go1.24.4 linux/amd64. For JSON parsing, use the standard library.

- The Ruby execution environment is Ruby 3.2.2 x86\_64-linux. For JSON parsing, use the standard library json module.

- The Rust execution environment is based on Ubuntu 20.04, using rustc 1.75.0 for compiling Rust code and cargo 1.75.0 for package management and build automation. For JSON parsing, use the serde\_json package.

- The Haskell execution environment is the Glorious Glasgow Haskell Compilation System (GHC) version 8.6.5. For JSON parsing, use the aeson package.

- The Perl execution environment is Perl version 5.30.0. For JSON parsing, use the JSON module from the libjson-perl package.

As a helpful AI Assistant, you provide well-reasoned and detailed responses by first thinking through the reasoning process as an internal monologue and then presenting the user with the answer. Provide your response in the following structured format: <think>
{...}
</think>
<answer>
\verb|```|\{language\}
\{code\}
\verb|```|
</answer>. In the section enclosed by <answer> and </answer> tags, ensure that only the transpiled code is included in accordance with the given format, such as \verb|```|python
{code}
\verb|```|.

[User]
Transpile the provided \{source language\} implementation into a functionally equivalent implementation in \{target language\}:

\verb|```|\{source language\}
\{source program\}
\verb|```|

\end{tcolorbox}

    \caption{Prompt template used for the transpilation task in both training and evaluation. The system prompt defines the task and the LLM’s role, specifies the detailed target code-execution runtime environments, and prescribes the required output format. The user prompt provides explicit instructions to transpile a given source program.
    }
    \label{fig:prompt-example}
\end{figure}